\documentclass{article}
\usepackage{arxiv}

\usepackage[utf8]{inputenc} 
\usepackage[T1]{fontenc}    
\usepackage{hyperref}       
\usepackage{url}            
\usepackage{datetime}       
\usepackage{booktabs}       
\usepackage{amsfonts}       
\usepackage{nicefrac}       
\usepackage{microtype}      
\usepackage{graphicx}
\usepackage[style=authoryear,backend=bibtex]{biblatex}
\usepackage{doi}
\usepackage{xcolor}

\usepackage{amsmath}

\hypersetup{colorlinks = true,
linkcolor = purple,
urlcolor  = blue,
citecolor = cyan,
anchorcolor = black}

\title{Using machine learning to inform harvest control rule design in complex fishery settings}

\newdate{articleDate}{9}{7}{2025}
\date{\displaydate{articleDate}}

\makeatletter
\let\@fnsymbol\@arabic
\makeatother

\author{Felipe Montealegre-Mora\\
Eric and Wendy Schmidt Center for Data Science and Environment\\ University of California,  Berkeley\\Department of Environmental Science, Policy, and Management\\ University of California,  Berkeley\\\AND
Carl Boettiger\\
Department of Environmental Science, Policy, and Management\\ University of California,  Berkeley\\\AND
Carl J. Walters\\
Institute for the Oceans and Fisheries,\\ University of British Columbia,\\\AND
Christopher L. Cahill\\
Quantitative Fisheries Center, Department of Fisheries and Wildlife,\\ Michigan State University,\\}

\renewcommand{\headeright}{}
\renewcommand{\undertitle}{}
\renewcommand{\shorttitle}{}

\hypersetup{
pdftitle={\@title},
pdfsubject={},
pdfauthor={\@author},
pdfkeywords={Machine learning,Reinforcement learning,Fishery science,Walleye,Spasmodic fishery},
addtopdfcreator={Written in Curvenote}
}

\bibliography{main.bib}
\begin{document}
\maketitle

\begin{abstract}
In fishery science, harvest management of size-structured stochastic populations is a long-standing and  difficult problem.
Rectilinear precautionary policies based on biomass and harvesting reference points now represent a standard approach to this problem.
While these standard feedback policies are based on analytical or dynamic programming solutions assuming relatively simple ecological dynamics, they are often applied to more complicated ecological settings in the real world.
In this paper we explore the problem of designing harvest control rules for partially observed, age-structured, spasmodic fish populations using tools from reinforcement learning (RL) and Bayesian optimization.
Our focus is on the case of Walleye fisheries in Alberta, Canada, whose populations display variable recruitment dynamics.
We optimized and evaluated policies using several complementary performance metrics representing key tradeoffs in harvest management.
The main questions we addressed were: 1. How do standard policies based on reference points perform relative to numerically optimized policies? 2. Can an observation of mean fish weight, in addition to stock biomass, aid in policy decisions?
\end{abstract}

\keywords{Machine learning, Reinforcement learning, Fishery science, Walleye, Spasmodic fishery}

%

\section{Introduction}

Over the past two decades a wealth of research has sought to solve sequential decision problems in diverse fields such as engineering, robotics, and control theory.
This work, collectively referred to as Reinforcement Learning (RL), has now advanced to the level needed to outperform human experts in many fields \cite{sutton, bertsekas2019reinforcement}.
For example, the application of RL has revolutionized games like chess, where the world's top chess engines now almost always defeat elite players.
Beyond outperforming experts, RL offers a new perspective on previously unsolvable problems---and in the case of chess, top players now regularly incorporate strategies discovered by engines into their repertoires.
Lessons like these seem relevant to fisheries scientists, as a number of sustainability problems lie at the intersection of age-structured population dynamics and sequential decision making under uncertainty \cite{walters1978ecological}, and for which traditional methods like dynamic programming or analytical approaches break down due to the ``curse of dimensionality.''
It is in these contexts that RL shows promise, and might serve as a useful guide for improving our intuition of feedback policy design.

In the absence of tractable analytical or dynamic programming solutions, simulation-based approaches like Management Strategy Evaluation (MSE) have been used to evaluate trade-offs among alternative policies or harvest control rules \cite{punt2016management}.
In MSE, analysts first specify policies to test a priori, and only after the specification of a policy set is simulation then used to quantify the relative performance of those policies against explicit objective(s).
While extensive application of the MSE approach has shown it to be a useful tool for informing harvest policy (e.g., \cite{edwards2016management}), it remains constrained to select among the a priori policy set chosen by analysts.
Phrased differently, if a particular policy is not included a priori due to a lack of creativity on behalf of analysts or perhaps due to the constraints of national legislation (e.g., see \cite{dfo2006}), then it simply is not possible to learn whether some alternative harvest control rule might outperform those tested.
In the context of feedback policy design this may be problematic, because feedback policy design is often counterintuitive, and because analysts typically limit themselves to 1-dimensional control rules based on stock biomass as per legislation \cite{punt2022framework, sainsbury2000design}.
For multidimensional population dynamics models (e.g., age-structured models, or community models), there is no theoretical guarantee that a single quantity (e.g. stock biomass) is sufficient to specify the optimal harvesting rate (see, e.g., discussion in \cite{walters2002stock}).

The dynamics of age-structured populations occur in high dimensional spaces, and thus in some situations it is possible for different states to correspond to the same total stock biomass.
For example, in a standard  age-structured model, many small fish can have the same total biomass as a few large fish.
Unsurprisingly, managers might prescribe different management actions  in  these different situations (see similar arguments in \cite{hilborn2002dark}).
Dimensionality problems such as this makes it difficult to specify policies to test \textit{a priori}.
While complexity due to age-structure makes it difficult to specify effective policies \textit{a priori}, it also implies that managers have access to more information than total biomass in some situations---this additional information can be useful for informing harvesting decisions.
For instance, knowing the mean fish weight in addition to total stock biomass could help the manager distinguish a population with many small fish from one with a population composed of mainly large fish.
Sources of additional information could be especially useful for managing fisheries with complex population dynamics.

Fish populations with highly variable or spasmodic recruitment are some of the most challenging systems to learn from and in which to inform fisheries management \cite{hjort1914fluctuations, caddy1983historical}, and thus serve as a useful test case in which to explore the utility of RL methodologies for improving feedback policy design.
In this paper we define highly variable or spasmodic recruitment to mean that a fish population exhibits infrequent large year classes of at least 10-50 times the long-term average recruitment level (see \cite{caddy1983historical}).
While by definition such events are rare in any one system, a review of the literature reveals that these types of fluctuations occur with some regularity in systes throughout the world (e.g., see \cite{fisch2019comparison}; \cite{licandeo2020management}).
For example, while effective fisheries management helped rebuild Northeast Atlantic fish stocks, it appears that of the stocks that displayed the strongest recoveries, record large year classes occurred at low abundance that drove stock productivity upward and out of low, collapsed states (see \cite{zimmermann2019improved}).
Similarly, Atlantic Redfish \textit{Sebastes fasciatus} stocks off the eastern coast of Canada exhibit spasmodic recruitment fluctuations that make the application of standard stock assessment techniques and MSE challenging \cite{licandeo2020management}.
In inland systems, Walleye \textit{Sander vitreus} in Lake Erie increased from low abundance due to a large recruitment event in 2003, and this cohort has continued to support the bulk of one of most economically valuable recreational fisheries in the world for nearly a decade \cite{schmitt2020does}.
Each of these large recruitment events affected both population status and the users who rely upon those populations.
However, such events are almost always written off as ``environmental effects'' and are notoriously difficult to predict with the reliability needed to inform feedback policy design (see \cite{myers1998environment}; \cite{punt2014fisheries}).

\begin{figure}[!htbp]
\centering
\includegraphics[width=0.875\linewidth]{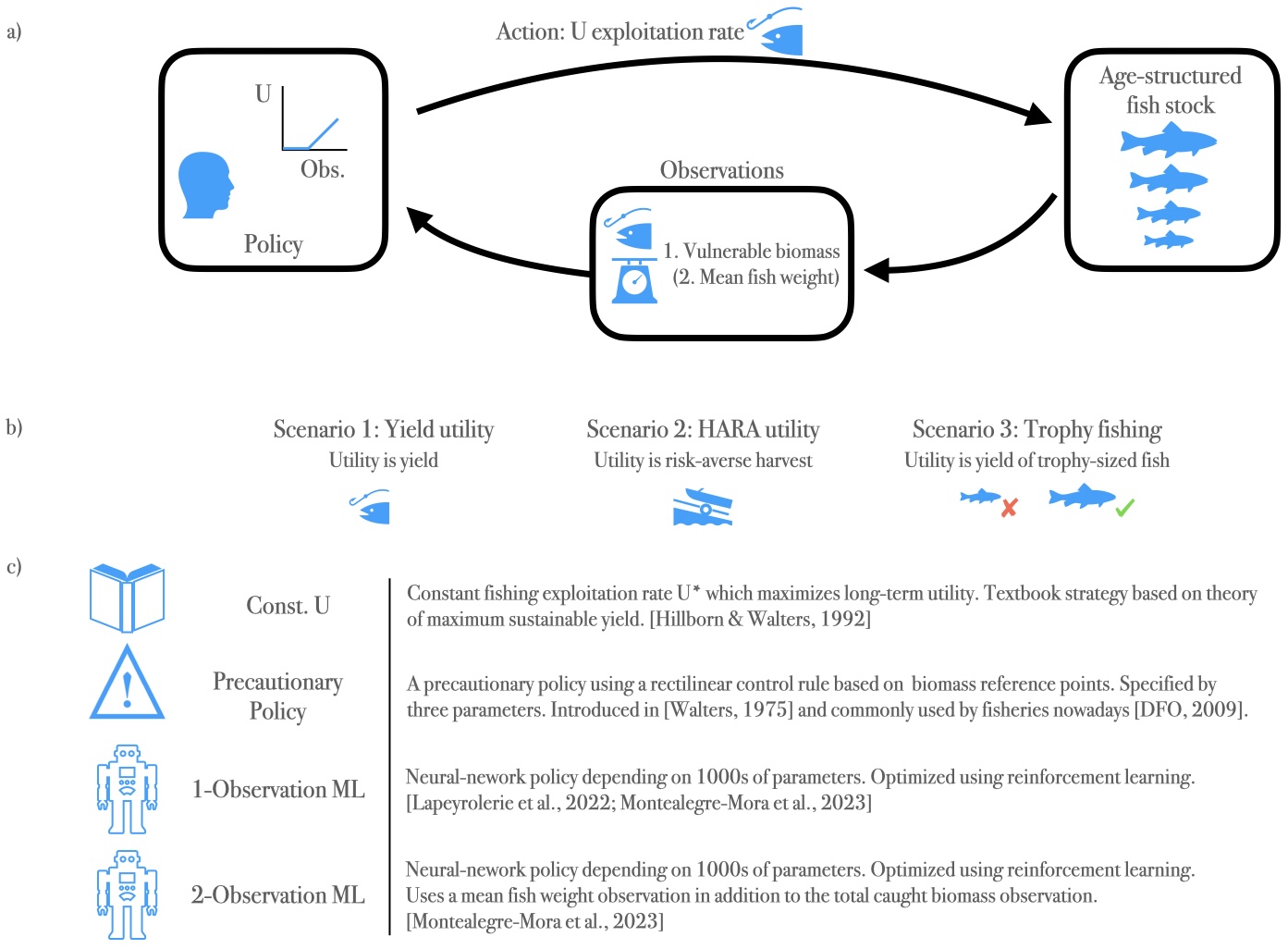}
\caption[]{Conceptual framework for our methodology. a) The dynamics of a Walleye population are simulated using a size-structured stochastic model with 20 size classes (right box).
The state of the population is specified by the biomass in each size class, i.e. the system's state is 20-dimensional.
Policy decisions are based on either one or two observations are gathered out of this 20-dimensional vector (middle box).
These observations are gathered through a simulated survey conducted by the fishery management.
Vulnerable biomass: the total biomass vulnerable to the survey gear, mean fish weight: the mean weight of fish vulnerable to the survey equipment.
A quantitative policy uses these observations to fix an instantaneous fishing exploitation rate (left box). b) We optimize policies in three scenarios with differing utility functions.
Total harvest: long-term yield. Hyperbolic additive risk-averse (HARA): a risk-averse utility function which penalizes inter-annual variability.
Trophy fishing: a utility function which only values large fish, while small fish do not contribute to the utility. c) Four types of policies are optimized. The first three policies-----constant-U, Precautionary Policy, and 1-Observation ML---use only the vulnerable biomass observation, whereas 2-Observation ML uses both observations.}
\label{fig:conceptual}
\end{figure}

Little work has examined the implications of large recruitment events on the performance of feedback policies in age-structured populations, even though feedback policies are considered the de facto standard for managing fisheries exploitation worldwide (see \cite{free2023harvest}; \cite{silvar2022empirical}; however, see \cite{licandeo2020management}).
This is noteworthy because nearly all of the theoretical work underpinning harvest control rule design has assumed populations exhibit uncorrelated recruitment anomalies originating from standard, stationary statistical distributions (see \cite{walters1975optimal}; \cite{walters1978ecological}; \cite{reed1979optimal}; however see \cite{parma1990experimental}; \cite{hawkshaw2015harvest}).
In this paper, we apply RL and Bayesian optimization to the problem of designing harvest control rules (HCRs) for partially-observed age-structured populations exhibiting highly variable recruitment dynamics (see Figure~\ref{fig:conceptual}a).
Specifically, we use RL methodologies to explore whether multi-dimensional control rules-----particularly rules depending on the total stock biomass and mean fish weight-----might be helpful for managing age-structured, spasmodically recruiting populations.
We focus our case study on a recreational Walleye fishery managed via harvest lottery in Alberta, Canada (see \cite{sullivan2003active}), as recent work showed these populations recovered from collapse due in part to large positive recruitment anomalies \cite{post2002canada, cahill2022unveiling}.
We also compare the policies obtained by numerical optimization with a rectilinear `default' precautionary rule recommended by the government of Canada \cite{dfo2006}.

In this study we evaluate HCRs with three types of utility functions: total harvest (i.e., yield maximizing), a risk-averse utility that prioritizes inter-annual consistency in catch, and a trophy fishing utility in which only sufficiently large fish are valued by harvesters (see Figure~\ref{fig:conceptual}b).
We optimize and evaluate four classes of HCRs: 1) constant exploitation rate ($U^*$), 2) a rectilinear precautionary rule (see [\cite{dfo2009}; Fig. 1] for a visualization) derived from $U^*$ and the average biomass $B^*$ to which the optimal $U^*$ policy converges to, 3) an unconstrained optimum rectilinear precautionary rule and 4) an HCR parametrized by a deep neural network using RL (see Figure~\ref{fig:conceptual}c). Our results show that considerable gains can be achieved in performance by optimizing policy parameters using RL and Bayesian optimization methodologies.
Whether mean weight was a useful variable for policy decisions depended on the choice of utility function.
Specifically, we found that mean weight was useful in the trophy fishing setting and, surprisingly, not when managers were interested in maximizing yield or risk-averse utility.

\section{Methods}

A collection of approximately 30 walleye lakes in Alberta are managed using a Special Harvest License (SHL), which enables managers to assign a Total Allowable Catch (TAC) to limit harvest, and which is unique for inland recreational fisheries in North America.
Presently, managers use an indicator-based approach to set TACs in any particular year based on standardized gill-netting surveys occuring in fall.
However, a recent study showed that harvestable surplus existed in many systems even though many management plans stated the goal of seeking to harvest for maximum sustainable yield (MSY) \cite{cahill2022unveiling}.
Here we aim to improve the scientific defensibility of this harvest allocation process in the Alberta SHL walleye program.
To do so, we extend a standard age-structured population dynamics model of walleye described in detail in \cite{cahill2022unveiling} and use it to simulate realistic population dynamics.
Briefly, we model population processes such growth in numbers at age through time as a function of Beverton-Holt stock-recruitment (see below), von Bertalanffy somatic growth in length-at-age (see \cite{cahill2022unveiling}), and total mortality as an additive process assuming total instantaneous mortality Z is equal to instantaneous natural mortality M plus fishing mortality rates imposed by recreational harvesters and which is modulated by vulnerability at age.
Unless stated otherwise parameter values for all relationships are drawn from average values estimated or used in \cite{cahill2022unveiling}.

\subsection{Population dynamics}

A central finding of \cite{cahill2022unveiling} was that recruitment dynamics were highly variable and spasmodic (see also \cite{caddy1983historical}).
Thus, we model a Walleye fishery population using a discrete-time, age-structured stochastic model with 20 age classes $(N_1,\dots,N_{a_{max}=20})$ with units of $\text{N}\cdot\text{ha}^{ -1}$.
Recruitment was modeled via the Beverton-Holt equation
\begin{equation}
\label{eq:popdyn}
N_{a, t+1} = 
\begin{cases}
  \frac{\alpha\text{SSB}_t}{1 + \beta \text{SSB}_t} r_t, & a=1,\\
  s_{a-1}\Big(1 - U_t v^{\text{harv.}}_{a-1}\Big) N_{a-1, t}, & 2\leq a < a_{max},\\
  s_{a_{max}-1} \Big(
    1 - U_t v^{\text{harv.}}_{a_{max}-1}
  \Big) N_{a_{max}-1} + s_{a_max} \Big(
    1 - U_t v^{\text{harv.}}_{a_{max}}
  \Big) N_{a_{max}}, & a=a_{max},
\end{cases}
\end{equation}
with a finite exploitation rate $U_t \in [0,1]$, and a \textit{spawning stock biomass}
\begin{equation}
\text{SSB}_t = \sum_a m_a W_a N_{a,t}
\end{equation}
where $W_a$ is the weight at-age and
\begin{equation}
m_a = \frac1{1+\exp\left(-0.5(a-a_{hm})\right)}
\end{equation}
is the maturity at-age, with $a_{hm}=6$.
This way, the total mortality rate at-age on year $t$ is
\begin{equation}
A_{a,t} = 1 - s_{a-1}\Big(1 - U_t v^{\text{harv.}}_{a}\Big).
\end{equation}
The parameters $\alpha$ and $\beta$ describe the juvenile survival as a function of $SSB_t$.
In (\ref{eq:popdyn}), $r_t$ is random deviate describing the spasmodic recruitment patterns observed in \cite{cahill2022unveiling}; it is distributed as
\begin{equation}
r_t \sim \begin{cases}
  \text{log-norm}(\mu=0,\sigma=0.4), & \text{with probability}\ 1-p_{big},\\
  \text{unif}(\text{min}=10, \text{max}=30), & \text{with probability}\ p_{big},
\end{cases}
\end{equation}
where $p_{big}=0.025$ is the probability of a large recruitment pulse.
The harvest vulnerability at-age is given by
\begin{equation}
\label{eq:vharv}
v_a^{\text{harv.}} = \frac{1}{1 + e^{-2(a-a_{hv})}}\, ,
\end{equation}
with an age at half-vulnerability $a_{hv}=5$ (see \cite{cahill2022unveiling} for more details).
All biomass quantities in this paper, including $SSB_t$, are in units of $\text{kg} \cdot \text{ha}^{ -1}$.

Large recruitment events are rare in any one fishery, however their occurrence happens at a rate much higher than would be predicted by the log-normal distribution alone \cite{cahill2022unveiling}.
Our model for $r_t$ is a minimalistic description of this dynamic which has explicit control over the rate of large recruitment events.
One can qualitatively see concordance between the biomass observation time-series obtained using our model for $r_t$ (see, e.g., Figure~\ref{fig:eps-um1}) and patterns found in the literature (for example, see Fig. 9 in \cite{cahill2022unveiling}).

Simulations were run for 1000 years in an attempt to capture the long-term effects of HCRs on population dynamics and performance criteria.
Specifically, the expected number of large recruitment years (i.e. ``successful'' Bernoulli trials for $r_t$) over a period of 1000 years was 25, which was judged to be high enough to capture the dynamics arising from a particular HCR.
Using nomenclature from the RL literature, we also refer to these 1000 year simulations as \textit{episodes}.

\subsection{Observations}

\begin{figure}[!htbp]
\centering
\includegraphics[width=0.625\linewidth]{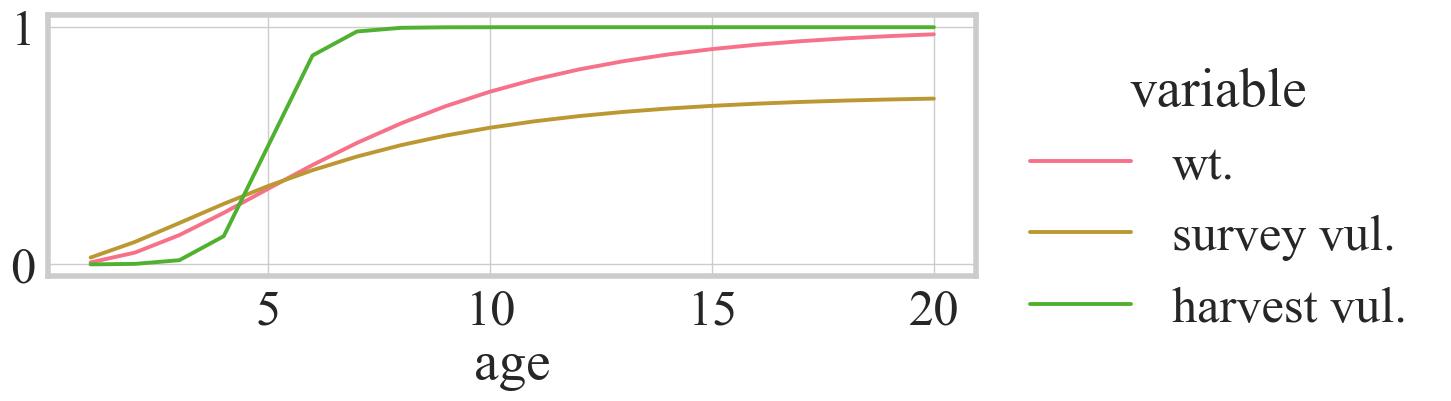}
\caption[]{Weight, survey vulnerability, and harvest vulnerability at-age.}
\label{fig:at-age}
\end{figure}

We model the observation process by simulating a gillnetting survey carried out by a management agency tasked with monitoring and managing the fishery.
These observations are then subsequently used by the HCR to set a fishing exploitation rate for each year.
We considered two types of observations.
The first is an estimate of the stock biomass vulnerable to the management agency's survey gear,
\begin{align}
  B_{survey} = \sum_{a=1}^{20}
  V_a^{\text{surv.}} W_a N_a,
\end{align}
where $V_a^{\text{surv.}}$ is the vulnerability-at-age of the survey.
We model $V_a^{\text{surv.}}$ as increasing with fish length, and use a von Bertalanffy function to describe length-at-age,
\begin{align}
V_a^{\text{surv.}} \propto (1 - e^{\kappa a})^\varphi,
\end{align}
with $\kappa=0.23$ and $\varphi=2.02$ (Figure~\ref{fig:at-age}, see \cite{cahill2022unveiling} for more details).
A second observation we consider is the \textit{mean weight} of fish in the survey,
\begin{equation}
\bar{W}_{survey} = B_{survey} / N_{survey},
\end{equation}
with
\begin{equation}
N_{survey} = \sum_{a=1}^{20} N_a V_a^{\text{surv.}}.
\end{equation}
Mean weight is easy for managers to observe and is important to consider in the context of spasmodically recruiting populations, since large recruitment events are correlated with dips in the mean weight of fish in the population.

We modeled observations as imperfect, with a multiplicative Gaussian observation error.
That is, policies do not observe $(B_{survey},\, \bar{W}_{survey})$ but rather, $(e_B B_{survey},\, e_W \bar{W}_{survey})$, where $e_B$ and $e_W$ are randomly sampled each year according to
\begin{equation}
e_B, \, e_W \sim \text{Normal}(\mu=1,\, \sigma=0.1).
\end{equation}

Here we emphasize that the system is thus only partially observed \cite{memarzadeh2019resolving}:
while the system dynamics unfold in the high-dimensional space defined by the biomass of each age class, most harvest control rules used to manage fisheries only observe some total measurement of stock biomass (with error) and apply a recommended Total Allowable Catch (TAC) based on this number.
It is also worth noting that while the model dynamics are Markovian in the full state space, the dynamics of these two observed states are not, making this a so-called Partially Observed Markov Decision Process, or POMDP.
This real-world problem that surveys provide imperfect information on overall abundance leads to a mathematical inconvenience that significantly increases the technical difficulty  of finding an optimal solution using classical tools like dynamic programming, and, as a result, the analysis of ecological POMDP problems has been restricted to models that make simpler ecological assumptions  (e.g., see \cite{williams2022partial}).

\subsection{Utility models}

We consider three utility models.

\begin{enumerate}
\item \textit{Yield maximizing utility:} the utility in each year is given by the total harvested biomass at that year,
\begin{equation}
\text{Utility}_{yield}(t) = \sum_{a=1}^{a_{max}} W_a U_t v^{\text{harv.}}_a.
\end{equation}

\item \textit{Risk averse utility}: a hyperbolic additive risk-averse utility function (also known as `HARA utility') which values inter-annual stability in harvests,
\begin{equation}
\text{Utility}_{\text{HARA}}(t) = \text{Utility}_{yield}(t)^{\gamma},
\end{equation}

where the parameter $\gamma=0.6$ specifies the risk aversion (see, e.g., \cite{collie2021harvest}).
Notice that the power of $\gamma$ attenuates the utility derived from large harvests.
Thus, low-and-stable harvests can perform relatively better with respect to $\text{Utility}_{\text{HARA}}$ than large-but-sparse harvest peaks.
We note that this is similar to using the natural log of catches as is commonly done in MSE.
Here the specific risk aversion $\gamma$ is set to a specific value for clarity, and we note that this parameterization converges on total harvest utility as  $\gamma\rightarrow1.0$.
Notice that if $\text{Utility}_{\text{yield}}(t) < 1$, then $\text{Utility}_{\text{yield}}(t) < \text{Utility}_{\text{HARA}}(t)$ and thus one can expect HARA utility values to be numerically larger than yield.
\item \textit{Trophy fishing utility}: here, harvested fish only contribute to the utility if they are above a certain size (age) class, $a > a_{thr.}$,
\begin{equation}
\text{Utility}_{trophy}(t) = \sum_{a>a_{thr.}} W_a U_t v^{\text{harv.}}_a,
\end{equation}

and we set $a_{thr.}=10$.
\end{enumerate}

We optimize the performance of policies with respect to the total utility obtained over an episode, i.e.,
\begin{equation}
\text{Utility} = \sum_t \text{Utility}(t).
\end{equation}

We considered utility functions one and two because they represent commonly acknowledged goals of fisheries management, that is maximizing yield and stabilizing harvests, which are at odds with one another and represent high harvest rates and high interannual variability in catches (maximizing yield) or low, but consistent harvest rates (maximizing risk-averse utility; see also \cite{walters1996fixed}; \cite{collie2021harvest}).
Moreover we included the trophy fishing function to explore how our analysis would change for more complex, size-dependent, utility functions such as those that may be needed in complex fisheries management.
Size dependence can be particularly relevant in cases where machinery to process harvests only operates within certain ranges of fish sizes, or perhaps when anglers only desire to retain large trophy-sized fish rather than valuing fish of any size equally (e.g., see \cite{murphy1996fisheries}; \cite{licandeo2020management}).

Within MSE, it is common for complex performance metrics containing different competing objectives to be used.
Defining the performance metrics used in MSE usually involves an active involvement of the spectrum of stakeholders.
This process was outside the scope of the present study, and instead we derived our results in a series of simple utility models that express well-known harvest management trade-offs.
As such, our results provide a useful guide to understand the optimization problem with respect to more complex objectives.

\subsection{Harvest control strategies}

Here we describe the three classes of HCRs we consider in this paper.
We refer to them interchangeably as \textit{control rules}, \textit{rules} or \textit{policies} throughout.

\textit{\textbf{Optimal constant-$U$ (Const-U) policy.}} In this strategy, the agent applies the same finite exploitation rate $U^*$ rate each year, where $U^*$ is the exploitation rate that leads to the highest average utility \cite{hilborn1992harvest}.
In the case where utility is equal to yield, the optimal constant exploitation rate is equal to the MSY fishing exploitation rate, $U_{MSY}$.

\textit{\textbf{Precautionary policies.}} A piece-wise linear HCR determined by three parameters: two stock biomass reference points $X_1$, $X_2$ along the $x$-axis, and the fishing exploitation rate at high stock biomass, $Y_2$. The HCR is given by the following equations:
\begin{equation}
U_t = 
\begin{cases}
  0, & B_{survey}(t) < X_1,\\
  \frac{B_{survey}(t) - X_1}{X_2 - X_1}\times Y_2, & X_1 \leq B_{survey}(t) \leq X_2,\\
  Y_2, & \text{for } B_{survey}(t) > X_2.
\end{cases}
\end{equation}
For a visual guide of this policy, see \cite{dfo2009}, Fig. 1.
The parameters of this policy are often fixed using the optimal constant exploitation rate $U^*$ to define $Y_2=U^*$ (at high stock biomasses), and two reference points for stock biomass \cite{dfo2009}.
Here, we use reference points defined by $X_1=0.4 B^*$ and $X_2=0.8 B^*$, where $B^*$ is a reference biomass given by
\begin{equation}
\label{eq:equil}
B^* = \overline{\mathrm{yield}}(U^*) / U^*,
\end{equation}
where $\overline{\mathrm{yield}}(U^*)$ is the average yearly yield obtained by the constant exploitation rate policy $U=U^*$.


In contrast to cPP, we refer to the precautionary policy whose parameters $(X_1,\, X_2,\, Y_2)$ have been numerically optimized as the \textit{optimized precautionary policy (oPP)}.
We evaluate both of these HCRs in relation to our chosen utility functions.

We collectively refer to the three policy types described above as \textit{fixed policies}, in the sense that their functional form is fixed \textit{a priori} before the optimization begins.
The process of optimizing these fixed policies is loosely inspired by MSE, where several competing policies are evaluated using dynamic simulations \cite{punt2014fisheries}.
However, with the exception of the cPP our approach differs from MSE in that we explicitly search across a continuous  space of potential parameter values using Bayesian optimization procedures (see below).

\textit{\textbf{Reinforcement learning policies.}} This policy uses a neural network to express the harvest control rule.
We explore two cases: one where only the stock biomass observation is used, and one where an additional mean weight observation is used.
Mathematically,
\begin{equation}
U_t = f_\theta(\text{Obs}_t),
\end{equation}
where $f_\theta$ is a neural network with parameters $\theta$, and $\text{Obs}_t$ are the observations obtained at year $t$.

We optimized two different scenarios: one single-observation RL policy (\textit{1RL}) in which only the biomass observation is used,
\begin{equation}
\text{Obs}_t = B_{survey}(t),
\end{equation}
and a two-observation RL policy (\textit{2RL}) in which biomass and mean weight were both used
\begin{equation}
\text{Obs}_t = (B_{survey}(t),\, \bar{W}_{survey}(t)).
\end{equation}

In the 1-observation scenario, we used a 3-layer feed-forward network with layer sizes [64, 32, 16],
while in the 2-observation scenario, we used layer sizes of [256, 64, 16].
We experimented using a [64,32,16] feed-forward network for the two-observation case, as well as other network geometries including thinner networks, wider networks, and deeper networks with 4 or 5 layers.
Among the geometries we tested, we found that all policies either performed equally well or worse (in terms of average utility) to the geometries we present here.
Because of this, we will not describe in detail these explorations.\footnote{This paper's companion open-source code at \href{https://github.com/boettiger-lab/rl4fisheries}{https://github.com/boettiger-lab/rl4fisheries} facilitates this exploration for the interested reader.}

\subsection{Bayesian optimization of ``fixed policy'' controls}

We used the framework of Bayesian optimization to find optimal parameters for the fixed policy HCRs.
A general introduction to this type of algorithm may be found on \cite{frazier2018tutorial, skopt}.
Bayesian optimization algorithms solve the problem of minimizing an unknown objective function  which one can only query with some stochasticity, i.e. where instead of being able to compute $f(x,\, y,\, z,\, \dots)$ exactly, we are only able to compute $f(x,\, y,\, z,\, \dots)+r$, where $r$ is random noise.
Thus, this type of algorithm is designed to be able to approximate a solution to the minimization problem
\begin{equation}
\mathrm{argmax}_{x,y,z,\dots}\, f(x,\, y,\, z,\, \dots)
\end{equation}
even when one can only evaluate $f$ ``imperfectly.''
These algorithms are most useful in scenarios where evaluating $f$ is computationally expensive because they tend to require much fewer function evaluations of $f$ than brute force optimization approaches.

In our case, the arguments $(x,\, y,\, z,\, \dots)$ are the parameters in the HCR function.
That is, for the precautionary policy, the arguments are $(X_1,\, X_2,\, Y_2)$, whereas for the $U_{MSY}$ policy, the only argument is the exploitation rate $U$. We maximize the average utility obtained by a policy by optimizing over these parameters. For example, for the constant exploitation rate policy,
\begin{equation}
\mathrm{argmax}_F\, \mathbb{E}[\text{Utility}(U)],
\end{equation}
where $\text{Utility}(U)$ is the utility obtained by simulating an episode using a constant $U$.
Note that $\text{Utility}(U)$ is thus a random variable since the dynamics are stochastic.

Because of our system's high stochasticity, evaluating the mean episode reward afforded by any policy requires taking the average across many episodes.
We heuristically used 250 episodes for this average.
The utility generated by any policy in any one episode within our model is driven by the population's productivity that episode, i.e.,
\begin{equation}
\pi = \sum_{t=1}^{1000} N_1(t).
\end{equation}
In Appendix B we show that the empirical cumulative distribution function for $\pi$ using 250 episodes has approximately converged.
This indicates that this choice of $N=250$ episodes can be expected to be sufficient to estimate the episode utility of a policy.\footnote{We additionally optimized, \textit{a posteriori}, fixed policies using $N=350$ and $N=450$ episodes and found essentially identical optimal utilities obtained. This further indicated that our choice of $N=250$ was sufficient for optimization.}

We used the Gaussian process minimizer algorithm from the \textit{scikit-optimize} Python package to perform this optimization, and allowed the algorithm to evaluate the objective function at 70 different points.
The number of points was chosen heuristically as it appeared high enough for the optimization to converge, but remained low enough to provide reasonable runtimes.
Further details of how this optimization procedure is performed may be found in the companion open-source code.\footnote{Found at \href{https://github.com/boettiger-lab/rl4fisheries/blob/main/scripts/tune.py}{scripts/tune.py}}

\subsection{Harvest control via neural network: reinforcement learning optimization}

In this section we give a small overview of the RL methods we used to optimize neural network policies.
For a broader introduction to RL, its methods, and applications in quantitative ecology, see \cite{lapeyrolerie2022deep}.
Reinforcement learning is a class of strategies to approach sequential decision problems in which a simulated \textit{``agent''} interacts with a simulated \textit{``environment''}.
The agent here represents the decision-maker, who observes and acts on the environment.
In the context of our paper, the agent is a manager tasked with making a quota decision each year, while the environment is an age-structured model of the fish population.
In RL, the agent can be restricted to only observe partial information about the state of the environment.
Here we consider imperfectly measured observations of the total stock biomass and the mean fish weight, but emphasize that additional observations could be passed to the agent.

The simulation is broken down into year-long time-steps.
At the beginning of each time-step, the agent prescribes an action that it applies to the environment.
Subsequently, the environment changes its internal state due to this action taken by the agent, and outputs an observation and a reward back to the agent.
The observation is used by the agent to take the next action.
A visualization of this is shown in \cite{lapeyrolerie2022deep}, Fig. 1.

In the context of this work, one can think of the manager prescribing a TAC as the RL agent making decisions given some partially observable fish population (or environment).
We call \textit{training} the process by which the agent optimizes the values of the neural network weights $\theta$ (i.e., the process by which the RL algorithm optimizes the HCR).
During training, the agent collects data on time-step interactions with the environment, slowly learning which actions to avoid given an observation, and which actions to encourage.
This is done, in broad terms, using a variation of stochastic gradient descent, which takes advantage of the fact that gradients of the reward space can be efficiently computed on neural networks using back propagation (for an introduction to back propagation, see \cite{rojas2013neural}.

A variety of algorithms exist for training RL agents, each with its strengths and drawbacks.
Here we focus on the Proximal Policy Optimization algorithm (PPO), which has been shown to perform well over a broad set of benchmark environments \cite{schulman2017proximal}.\footnote{We tested certain other RL algorithms, such as the \textit{Truncated Quantile Critics (TQC)} algorithm, as well. However, we found it hard to match the performance of PPO. Because of this, we do not include the analysis of policies obtained with these other algorithms here. The companion source code allows the user to easily reproduce our analysis for PPO and other RL algorithms.}
It moreover has also been shown to have a strong performance in problems related to population dynamics \cite{lapeyrolerie2022deep, montealegre2023pretty}.
As mentioned, the neural networks we use in our results are rather modest in size---with only a few thousand parameters.
The training times were also modest, comprising 6 million time-steps, or a bit under two hours of training on a commercial GPU.

\subsection{Policy evaluation}

After optimizing each HCR, we simulated $n=500$ episodes and recorded utility obtained by each policy in each episode.
We visualize this data in Figure~\ref{fig:rewards} where the (interpolated) density of utilities obtained by each policy is plotted, and we record summary statistics for this data in Table~\ref{tab:rew-table}.
Moreover, to get a more detailed comparison of the dynamics induced by each HCR, we simulated an additional episode where we recorded the stock biomass, mean fish weight and exploitation rate.
To improve comparisons between policies, we used the same time-series of stochastic deviations $\{r_t\}_{t=1, \dots, 1000}$ across all of the latter set of simulations.
Recall that we use a Beverton-Holt recruitment model in which $r_t$ is the random deviate for recruitment for year $t$ (see eq. (\ref{eq:popdyn})).
Finally, in order to compare policy responses in the aftermath of a large recruitment year, we performed $n=500$ simulations of short time-series (30 years) in which the first year was a large recruitment year, and the subsequent years had normal recruitment.
That is, for these simulations we used
\begin{align}
  r_1 &\sim \mathrm{Unif}(10,\, 30),\\
  r_t &\sim \mathrm{lognorm}(0,\, 0.4), \quad t>1.
\end{align}

\section{Results}

\begin{figure}[!htbp]
\centering
\includegraphics[width=0.9375\linewidth]{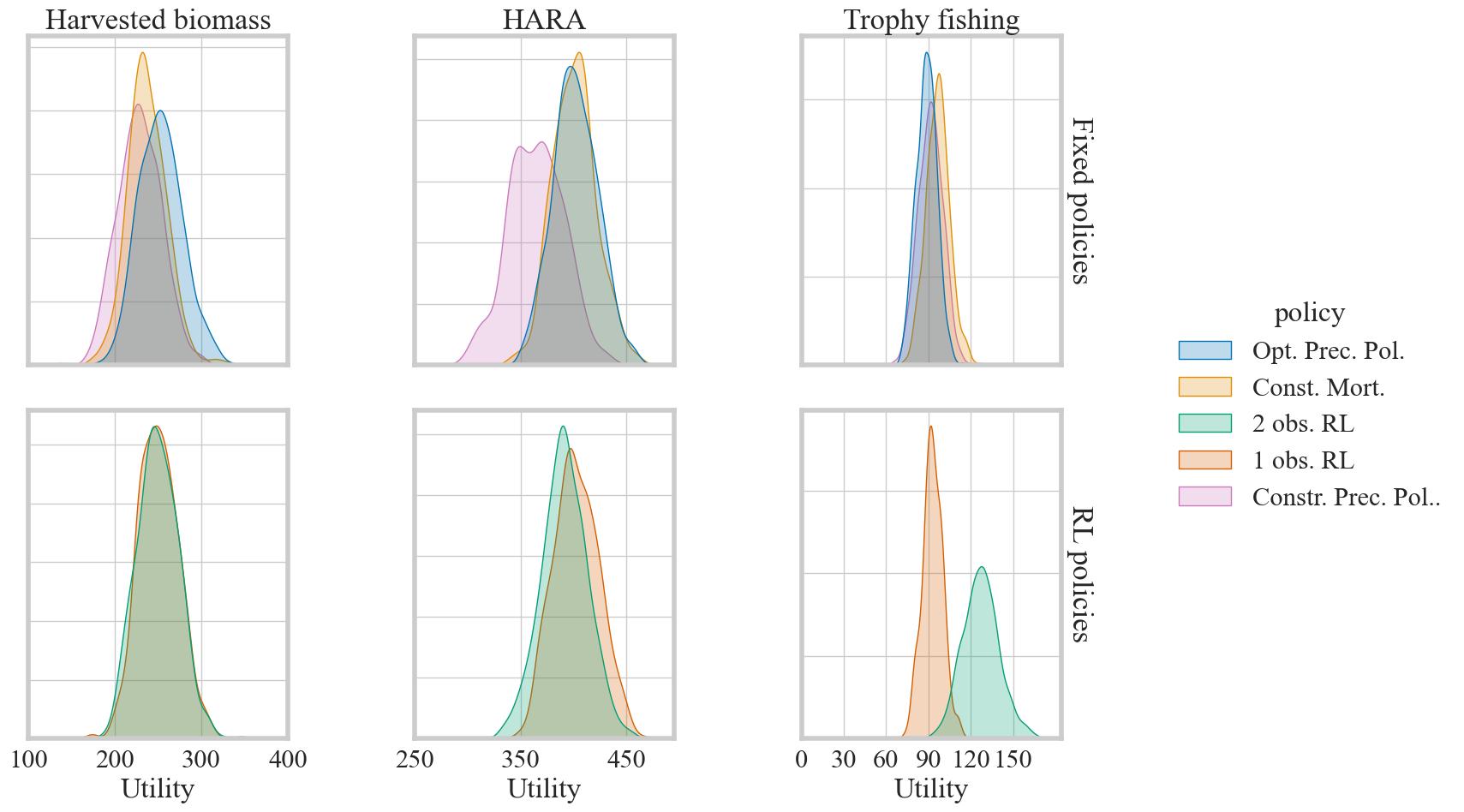}
\caption[]{Reward distributions for each of the five optimized policies in each of the three scenarios.
These distributions were interpolated from the total utility obtained in n=500 simulated episodes.
See associated Jupyter notebooks for details on this.}
\label{fig:rewards}
\end{figure}

In Table~\ref{tab:fixed-params} we show the parameter values obtained for the optimized fixed HCRs.
The episode utilities obtained by these HCRs are displayed in Table~\ref{tab:rew-table} and Figure~\ref{fig:rewards}.
With respect to the $\text{Utility}_{yield}$ and $\text{Utility}_{HARA}$ utilities, we find that nearly all policies obtain essentially equal amounts of utility----the only exception being the cPP policy which underperforms relative to the other policies in the HARA scenario.
In contrast to this, in the trophy fishing scenario (right column), we see that the 2RL control obtains about 30\% more utility than other harvesting policies.

\begin{figure}[!htbp]
\centering
\includegraphics[width=0.9375\linewidth]{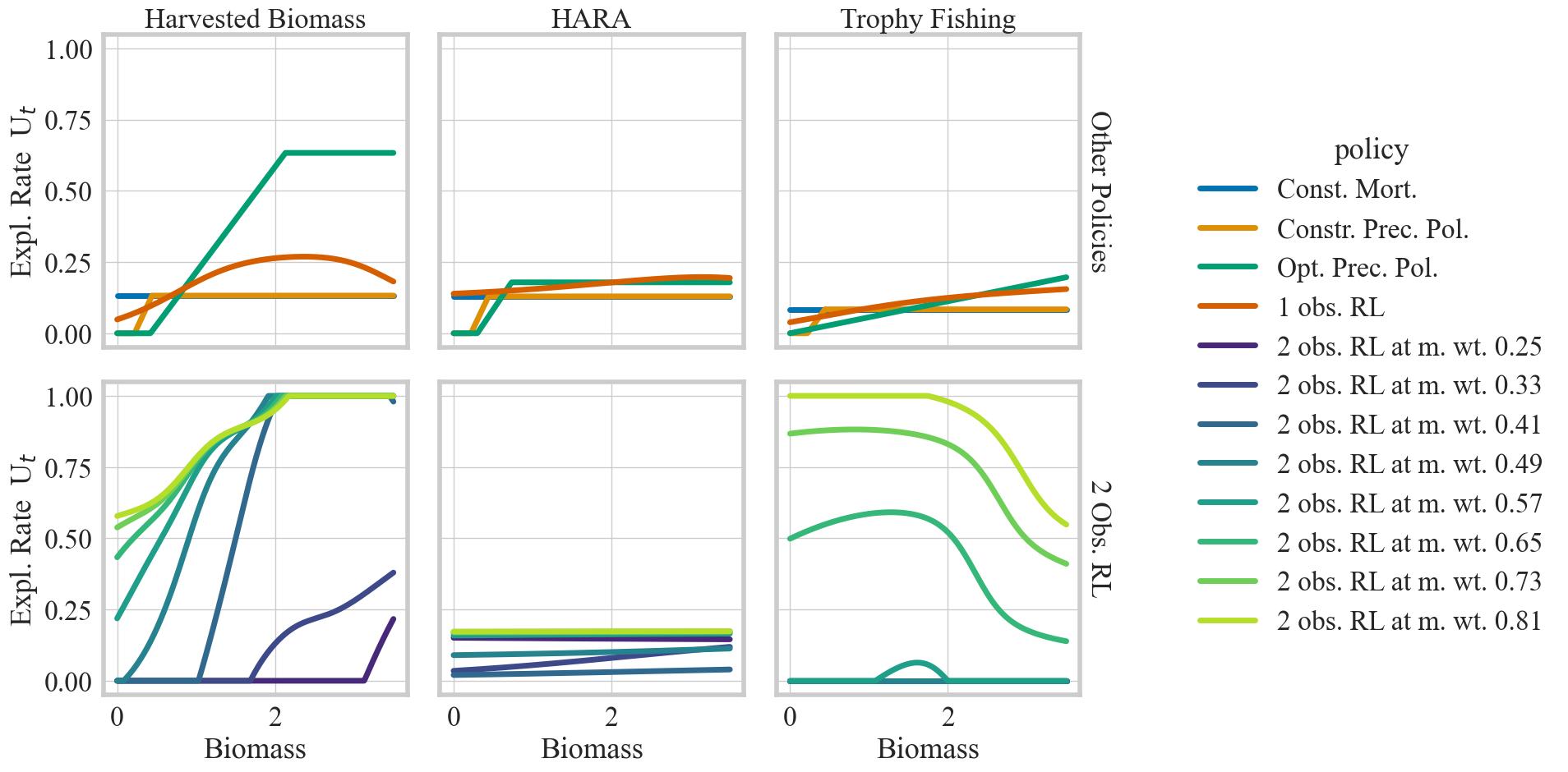}
\caption[]{Optimized policy functions for each of the scenarios.
Top row: policies which only use the vulnerable biomass observation to inform exploitation rate.
Bottom row: 2-Observation RL policy, where exploitation rate is informed by the vulnerable biomass observation as well as the mean fish weight observation.
The dependence of exploitation rate on mean weight is displayed with the color code (a gradient between green at high mean weight and violet at low mean weight).}
\label{fig:policies}
\end{figure}

The optimized HCRs are visualized in Figure~\ref{fig:policies}, where we plot exploitation rate as a function of stock biomass.
We discuss these plots in the following paragraphs.

\begin{table}
\centering
\caption[]{Optimal parameter values for fixed policies for each of the three utility models. For cPP we use $X_1=0.4 B^*$, $X_2=0.8 B^*$ and $Y_2=U^*$ as described in the methods section. As seen in the table, $U^*=0.132$, and by using Table 2 we find that $\overline{\mathrm{yield}}(U^*)\approx237/1000$. Thus, $B^*\approx1.813$.}
\label{tab:fixed-params}
\begin{tabular}{p{\dimexpr 0.250\linewidth-2\tabcolsep}p{\dimexpr 0.250\linewidth-2\tabcolsep}p{\dimexpr 0.250\linewidth-2\tabcolsep}p{\dimexpr 0.250\linewidth-2\tabcolsep}}
\toprule
\textbf{Policy} & \textbf{Yield} & \textbf{HARA} & \textbf{Trophy Fishing} \\
\hline
$U_{MSY}$ & $U=0.132$ & $U=0.129$ & $U=0.084$ \\
 & --- & --- & --- \\
 & $X_1=0.422$ & $X_1=0.301$ & $X_1=0.000$ \\
oPP & $X_2=2.132$ & $X_2=0.734$ & $X_2=5.751$ \\
 & $Y_2=0.633$ & $Y_2=0.179$ & $Y_2=0.323$ \\
 & --- & --- & --- \\
 & $X_1=0.725$ & $X_1=0.725$ & $X_1=0.725$ \\
cPP & $X_2=1.451$ & $X_2=1.451$ & $X_2=1.451$ \\
 & $Y_2=0.132$ & $Y_2=0.129$ & $Y_2=0.084$ \\
\bottomrule
\end{tabular}
\end{table}

\begin{table}
\centering
\caption[]{Summary statistics of the reward distributions for optimized policies. In each column, the highlighted utilities are within a standard deviation of the best-performing policy.}
\label{tab:rew-table}
\begin{tabular}{p{\dimexpr 0.250\linewidth-2\tabcolsep}p{\dimexpr 0.250\linewidth-2\tabcolsep}p{\dimexpr 0.250\linewidth-2\tabcolsep}p{\dimexpr 0.250\linewidth-2\tabcolsep}}
\toprule
Policy & Yield Util. & HARA Util. & Trophy Fishing Util. \\
\hline
oPP & \textit{252.82} +/-- 25.31 & \textit{401.73} +/-- 20.93 & 88.51   +/-- 7.17 \\
cPP & \textit{228.64} +/-- 25.03 & 364.55  +/-- 25.94 & 91.34   +/-- 8.76 \\
UMSY & \textit{237.28} +/-- 22.68 & \textit{400.58} +/-- 20.67 & 96.44   +/-- 8.24 \\
1RL & \textit{250.25} +/-- 22.82 & \textit{402.43} +/-- 21.14 & 92.73   +/-- 7.13 \\
2RL & \textit{249.47} +/-- 23.86 & \textit{391.27} +/-- 21.51 & \textit{126.90} +/-- 12.80 \\
\bottomrule
\end{tabular}
\end{table}

\begin{figure}[!htbp]
\centering
\includegraphics[width=0.9375\linewidth]{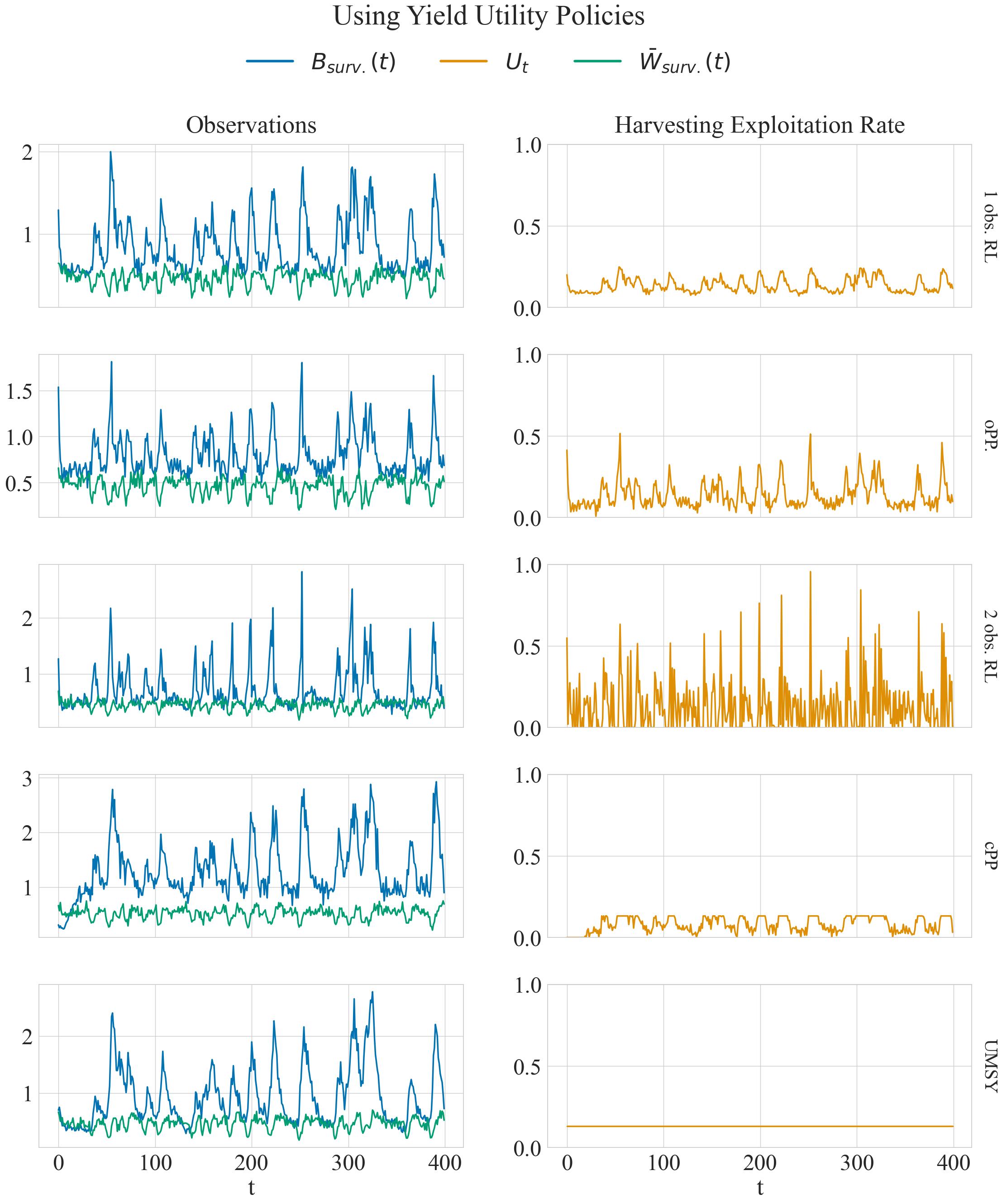}
\caption[]{First 400 years for an episode simulated with each of the HCRs optimized for \textit{Total harvest utility}.}
\label{fig:eps-um1}
\end{figure}

\textit{Yield utility (Figure~\ref{fig:policies}, left column)}.
In this scenario we observe a wide variety of policy shapes (Figure~\ref{fig:policies}) leading to very similar episode utility distributions (Figure~\ref{fig:rewards}, Table~\ref{tab:rew-table}).
Moreover, in Figure~\ref{fig:eps-um1} we observe that these different policies indeed lead to quite different dynamical patterns for the exploitation rate.
Our results suggest that there is a wide variety of control strategies available to a manager that lead to essentially equal behavior in long-term yield.
For example we observe may contrast the pulsed-fishing behavior of 2RL with UMSY, with the other policies' behavior lying somewhere in the spectrum between these two extremes.

\begin{figure}[!htbp]
\centering
\includegraphics[width=0.9375\linewidth]{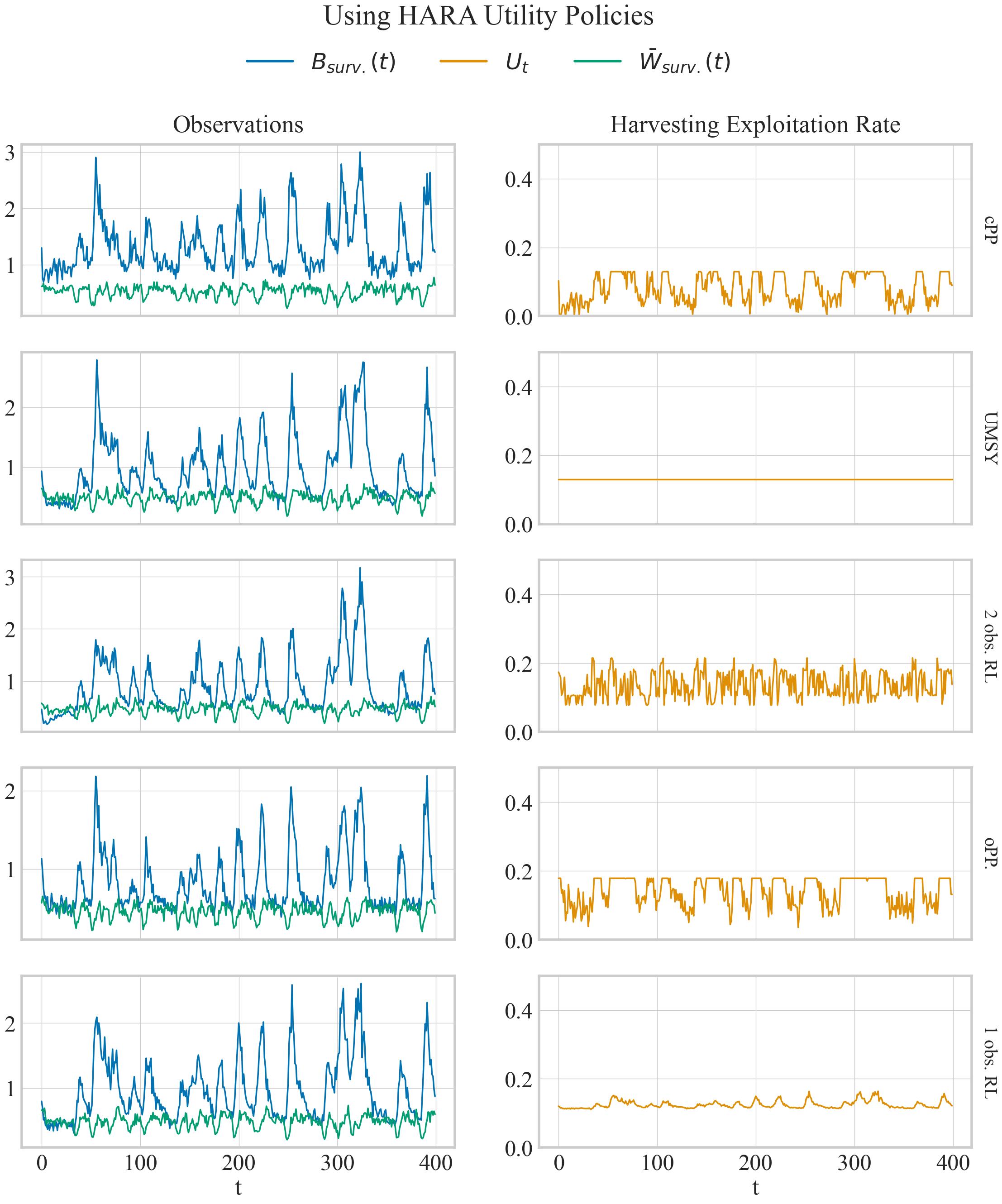}
\caption[]{First 400 years for an episode simulated with each of the HCRs optimized for \textit{HARA utility}.}
\label{fig:eps-um2}
\end{figure}

\textit{HARA utility (Figure~\ref{fig:policies}, middle column)}.
In this case, in contrast to the former scenario, we see that all policies converge to similar behavior to UMSY, with relatively flat $U$ curves as a function of observed stock biomass.
This behavior can be confirmed in Figure~\ref{fig:eps-um2}, in which we see that for all policies, $U_t$ seems to hover around the UMSY value $U\approx 0.12$ (Table~\ref{tab:fixed-params}).
We observe that the 2RL rule leads to noisy behavior which we believe is due to the observational noise coupled with the strong dependence of the policy on mean weight observations.
We believe that using a moving average window for mean weight observations, together with using larger networks and longer training times, could help to smooth this behavior.
However this exploration would be of limited interest due to the expectation that this way the 2RL policy would converge to UMSY.

\begin{figure}[!htbp]
\centering
\includegraphics[width=0.9375\linewidth]{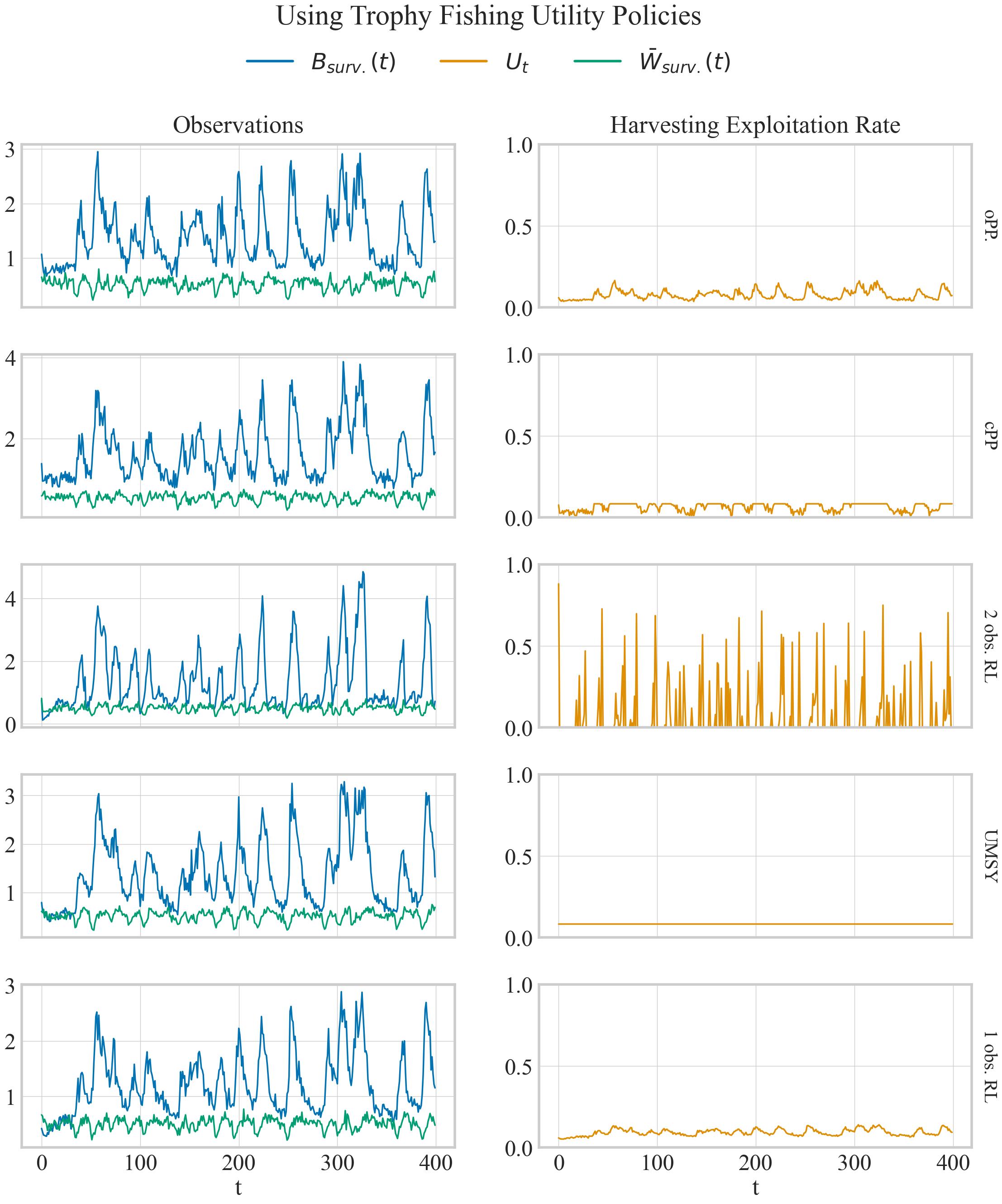}
\caption[]{First 400 years for an episode simulated with each of the HCRs optimized for \textit{Trophy fishing utility}.}
\label{fig:eps-um3}
\end{figure}

\begin{figure}[!htbp]
\centering
\includegraphics[width=0.9375\linewidth]{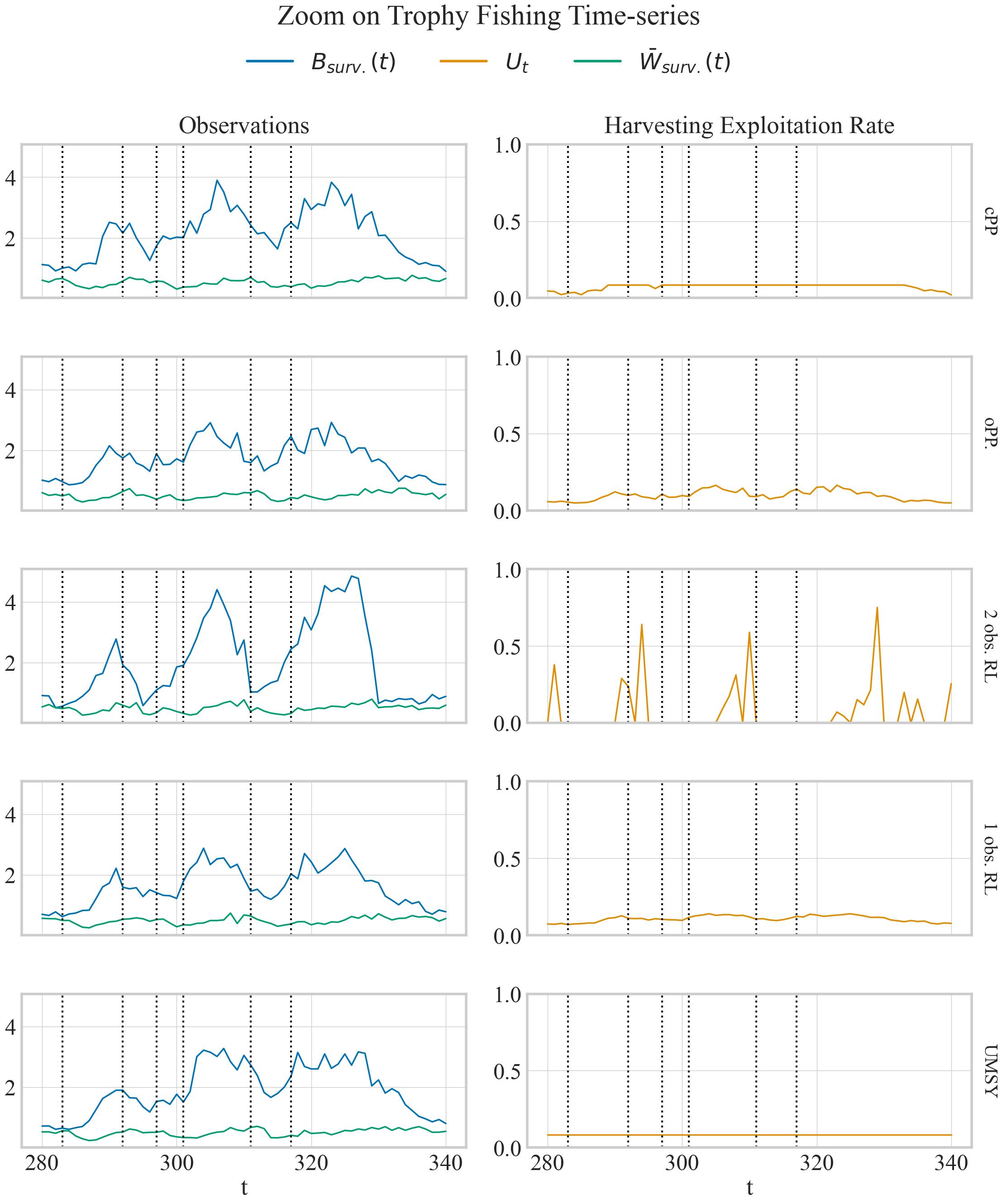}
\caption[]{A zoom into part of the episode for the trophy utility scenario.
We display large recruitment years as vertical dotted lines.}
\label{fig:eps-um3-zoom}
\end{figure}

\textit{Trophy fishing utility (Figure~\ref{fig:policies}, right column)}.
As previously pointed out, in this scenario we observe a marked advantage for 2RL with respect to all other policies in terms of utility obtained.
Examining Figure~\ref{fig:eps-um3} we may see that  2RL used a pulsed harvesting strategy, with short bursts of high exploitation rates, followed by periods of no harvests.
This behaviour can be contrasted with all other optimized policies we tested, which have exploitation rates that remain relatively stable over time (see Figure~\ref{fig:eps-um3}, right column).
While the increase of exploitation rate with respect to mean weight by the 2RL policy (Figure~\ref{fig:policies}) is clear, this policy has a highly non-intuitive property of decreasing exploitation with increasing biomass at high biomasses.
To understand the behavior of the 2RL policy, and its difference with respect to other policies we display a zoom into the same time-series in Figure~\ref{fig:eps-um3-zoom}, together with the times of large recruitment years.
Here we see that, by avoiding fishing in the years subsequent to a recruitment pulse, the 2RL agent is able to perform a large fishing pulse ($U\approx0.75$) on a population with high biomass and mean weight (with the class of the fishing pulse at $t\approx 320$ being about 10-15 years old at the time of the pulse).
In contrast, the other optimized policies have relatively stable fishing mortalities over time, with little response to large year classes.

\begin{figure}[!htbp]
\centering
\includegraphics[width=1\linewidth]{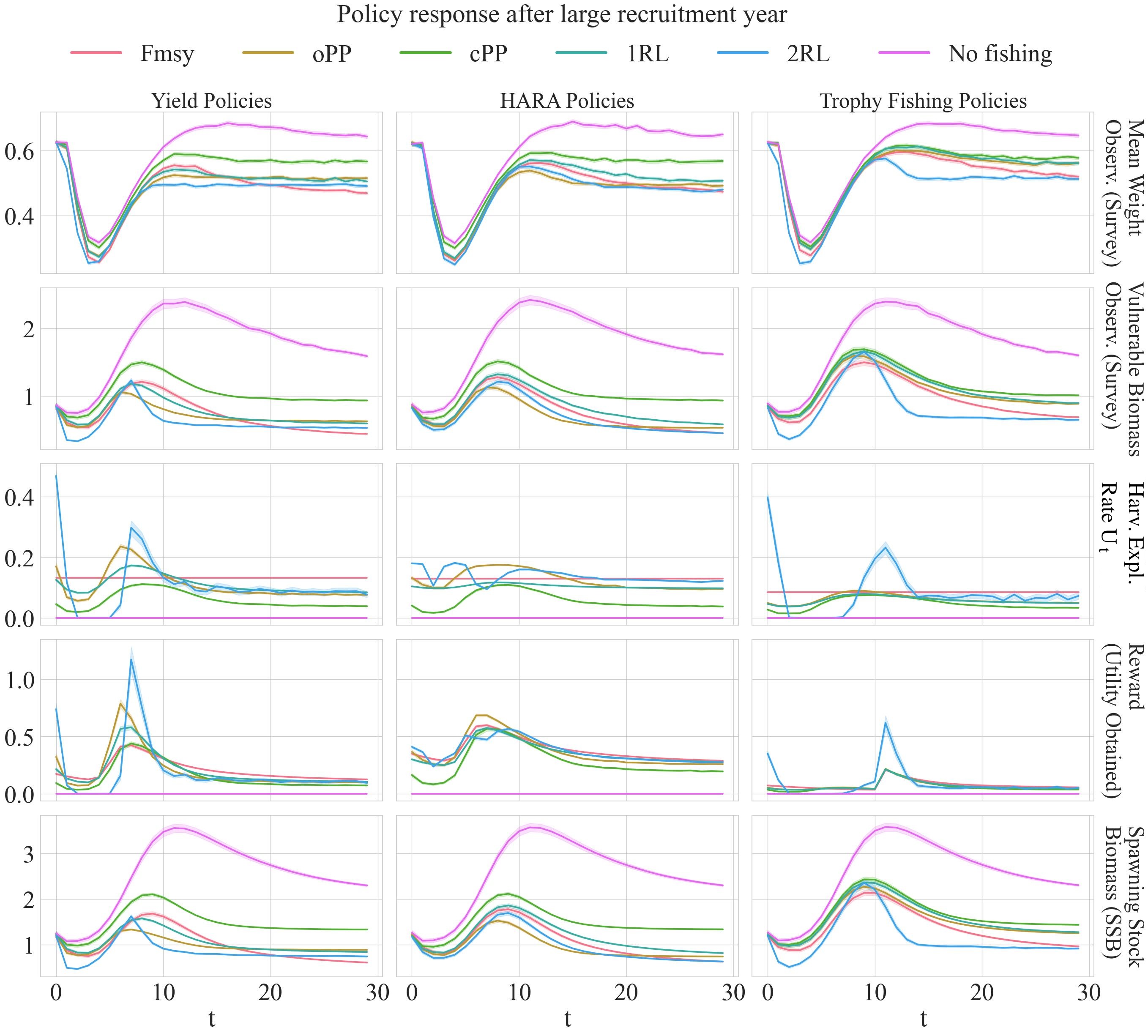}
\caption[]{Policy interactions with our dynamical model in the timesteps following a large year class.
For each optimized policy, we simulated 500 time-series of 30 years in which the first year is a large recruitment year and the subsequent years are normal recruitment years.
We plot the average response for each policy and shade the error bar region.
Additionally, we display the dynamics arising without fishing for comparison.}
\label{fig:aftermath}
\end{figure}

In Figure~\ref{fig:aftermath} we display the reponse of each optimized policy (in each of the utility scenarios) following a large recruitment year.
Immediately after the large recruitment year there is a dip in both the observed mean weight as well as the observed stock biomass.
The latter is due to the fact that survey vulnerability at-age is low for small age classes (Figure~\ref{fig:at-age}).
Similarly, the pronounced dip in mean weight at $t\approx5$ (rather than the dip being at $t=0$) is explained by the survey vulnerability at-age schedule.
In this plot, we observe that the 2RL fishing pulse in the trophy fishing scenario (right column) happens approximately between timesteps 8 and 15, around the age at which the large recruitment generation starts generating utility.

\section{Discussion}

Feedback policy design remains a difficult problem for resource management, particularly in the context of age- or size-structure, assessment errors in abundance estimates, and for fisheries with highly variable or spasmodic recruitment dynamics (e.g., see \cite{walters1986adaptive}; \cite{caddy1983historical}; \cite{walters2002stock}; \cite{williams2022partial}).
Using simulated data with these complexities, we have shown how RL can be used to inform or aid in harvest control rule design.
As such, our work represents an extension of early dynamic programming and analytical operations research used to find optimal policies for simple ecological models with explicit performance criteria (e.g., \cite{walters1975optimal}; \cite{walters1978ecological}; \cite{walters1996fixed}; \cite{reed1974stochastic}), and upon which nearly all of the precautionary harvest control rules used worldwide are currently based (see \cite{restrepo1999precautionary}; \cite{free2023harvest}).
In doing so, this work provides an additional computational framework that can be used to explore feedback policy design in apparently realistic ecological settings like those that simply break existing tools or approaches due to challenges arising from the ``curse of dimensionality'' \cite{woodward2005living, marescot2013complex, reimer2022structural}.

We find that, with respect to yield maximization, strikingly vast differences in the functional form of optimized policies have negligible effects in utility.
For example, in Figure~\ref{fig:aftermath} (left column) we can see that the different policies indeed have very different profiles of when fishing is performed (and thus, when utility is accrued) relative to a recruitment pulse.
In contrast to this behavior, in the middle columns of Figure~\ref{fig:aftermath} and Figure~\ref{fig:policies} we observe that all optimized policies (i.e. all policies except for cPP) lead to fairly similar fishing patterns and utility returns.
Thus, our results suggest that for this system there is a fair deal of flexibility in exploring performance metrics beyond yield, even in cases where maximizing yield is enshrined in the law (such as is the case for fisheries in the United States, \cite{act1996magnuson}).

We find the surprising result that mean weight observations mattered little in decision-making for two very different utility functions ($\text{Utility}_{yield}$ and $\text{Utility}_{HARA}$).
Our hypothesis at the start of the project was that this type of observation would lead to improvents in utility across the board, due to the fact that large recruitment events---and their associated effects on mean fish weight---are linked to the population dynamics of spasmodically recruiting fisheries.
This unexpected result speaks to the counterintuitive nature of feedback control (e.g., see \cite{moxnes2003uncertain}), particularly given the fact that dynamics of the system depends on the full state of the system rather than on only the stock biomass \cite{caddy1983historical, licandeo2020management}.
Consequently, it would be natural to expect that this would render stock biomass-based control rules ineffective when compared to policies which use further information about the age structure of the population for decision-making.
These findings show that many of the 1-D feedback policies discovered during the 1960s-1990s using dynamic programming and analytical methods (e.g., see \cite{walters1969generalized}, \cite{walters1975optimal}, \cite{walters1978ecological}, \cite{walters1996fixed}) were able to perform as well as the much more complicated policies we considered using RL.
This finding appears to reinforce a general result in the literature relating to the fundamental trade-off between policies that seek to maximize yield via constant escapement-like policies vs.
HARA utility-like policies that reduce the exploitation rate to low values and stabilize harvests through time (\cite{walters1996fixed}; \cite{collie2021harvest}).

In contrast to yield or HARA utility, the trophy utility policies that did not value small fish led to a dependence of the best fishing rate each year on stock biomass and the average size of fish, which essentially avoided harvest when a dominant cohort entered the population and drove mean size of the population down.
For this performance criterion, the best policies for maximizing yield of large or trophy sized fish waited until a strong cohort introduced by a recruitment pulse reached larger body size (i.e., until mean weight was higher).
This resulted in a time-dependent pulse fishing pattern reminiscent of bang-bang policies (see, e.g., \cite{walters1978ecological}), that was first shown to be optimal by \cite{walters1969generalized} when capturing and killing fish of all ages is unavoidable (see also \cite{botsford1985optimal}).
In Figure~\ref{fig:aftermath} (right column, third row), we see that on average 2RL policies produce a fishing pulse between timesteps 8-15 after the recruitment pulse, that is during the time-period in which the large cohort reaches an age or size that is valuable from a utility perspective.
After this, 2RL fishes slightly below the optimal constant exploitation rate on average, however, inspecting Figure~\ref{fig:eps-um3-zoom} we see that this average behavior may include smaller fishing spikes together with periods of no exploitation.
This behavior speaks to the tradeoff between total utility accrued on the one hand, and stability of utility over time on the other.
Given the extreme non-linearity of our trophy utility function---in which fish below a threshold age provide no utility---the tradeoff tilts the balance against stability (see, e.g., the utility curve in Figure~\ref{fig:aftermath}, right column, fouth row).
We velieve that our results point at a potentially more general pattern in which non-trivial age-dependence within the utility function makes age-structure observations (such as mean weight) valuable for the harvesting decision problem.
In this more general scenario the trade-off between stability and total utility can be expected to appear as well, albeit not necessarily with the same resulting tilt towards total utility.

In summary, the reinforcement learning simulated manager trained to maximize trophy utility, when given access to mean weight observations, monitors the age composition of the population in order to decide when to harvest.
Fishing pulses have been shown to be optimal within other situations \cite{walters1969generalized}, a behavior which is reproduced by our RL manager in the trophy fishing scenario.
In essence, the RL manager waits for large cohorts and adjusts the timing of its harvest to maximize utility from those large cohorts.

Our results for the trophy fishing metric also illustrate the scalability and generalizability of RL to new problems (see also \cite{sutton}), as these methods free the analyst from having to intuit the nuances of feedback control in specific situations with respect to specific objectives.
Specifically, while there is no standard functional form for HCRs dependent on biomass and mean fish weight, the flexibility of not requiring such a functional form enabled us to use RL to optimize policies in this scenario.
In this sense, RL policies strike a balance between being expressive (i.e. allowing for many different functional forms by tuning network parameters), and efficiency of optimizing neural network parameters (through backpropagation).
We believe such a tool may prove particularly useful when managers are interested in more complex objectives (e.g., \cite{hilborn1992objectives}; \cite{pascoe2017modelling}, \cite{salomon2023disrupting}; \cite{silver2022fish}; \cite{vaca2006analysis}) or in ecological settings in which there are no theoretical guarantees that the 1-D harvest control rules based on stock biomass remain effective (e.g., see discussion in \cite{walters2002stock}).
Moreover, recent literature has explored the importance of including complex environmental and multi-species interactions in feedback policy design (e.g. \cite{perryman2021review}), a context in which RL might provide novel insights.

\section{Future work}

In this paper, we developed an age-structured fish population dynamics model specifically designed for integration with reinforcement learning (RL) methodologies.
As interest in RL grows within the fisheries community \cite{ditria2022artificial, kuhn2025machine, ju2025model, montealegre2023pretty, nicolas2020deep}, our findings show that RL can identify and evaluate feedback policies in complex ecological settings that have resisted solutions via dynamic programming or analytical approaches.
This advance enables analysts to revisit longstanding challenges in harvesting theory that demand explicit treatment of partial observability, nonstationary ecological dynamics, and nonlinear trade-offs in utility functions---such as the trophy utility function examined here.
These contributions raise several unresolved but important questions:
Under what conditions can RL uncover ``pretty good'' policies---akin to Hilborn's concept of pretty good yield---that balance competing objectives?
How can RL be extended to support harvesting decisions in spatially structured systems, where local heterogeneity and connectivity complicate management?
And how might RL help design policies that reconcile performance with justice, especially in contexts where Indigenous strategies such as rotational harvests fall outside the scope of conventional management strategy evaluation?
These questions address technical challenges to harvest control theory and feedback policy design, particularly because optimization of feedback policies has historically been limited to low-dimensional policy functions that rely solely on stock biomass as the predictor for harvest.
We believe RL offers a flexible computational framework that makes these questions tractable and provides a path toward answering them.

\printbibliography

\end{document}